%% file: main.tex
\documentclass[10pt,twocolumn,letterpaper]{article}

\usepackage{cvpr}              
\usepackage{multirow}

\usepackage{colortbl}
\definecolor{lightgray}{rgb}{0.93,0.93,0.93}
\usepackage[pagebackref,breaklinks,colorlinks,citecolor=cvprblue]{hyperref}
\usepackage{pifont}
\usepackage{algorithm}  
\usepackage{algpseudocode}  
\usepackage{booktabs}
\usepackage{multirow}

\newcommand{\vct}[1]{\mathbf{#1}} 
\newcommand{\mat}[1]{\mathbf{#1}} 

\usepackage[misc]{ifsym}
\newcommand\blfootnote[1]{
    \begingroup
    \renewcommand\thefootnote{}\footnote{#1}
    \addtocounter{footnote}{-1}
    \endgroup
}
\usepackage[hang]{footmisc}

\definecolor{cvprblue}{rgb}{0.21,0.49,0.74}

\def\confName{CVPR}
\def\confYear{2025}

\title{Patient-Level Anatomy Meets Scanning-Level Physics: Personalized Federated Low-Dose CT Denoising Empowered by Large Language Model}

\author{
    Ziyuan Yang$^{1,2}$, Yingyu Chen$^{1,2}$, Zhiwen Wang$^{4}$, Hongming Shan$^{5}$, Yang Chen$^{6}$, Yi Zhang$^{2,3,*}$ \\
    $^1$College of Computer Science, Sichuan University, China \\
    $^2$Key Laboratory of Data Protection and Intelligent Management, \\Ministry of Education, Sichuan University, China\\
    $^3$School of Cyber Science and Engineering, Sichuan University, China\\
    $^4$School of Computer Science and Software Engineering, Southwest Petroleum University, China \\
    $^5$Institute of Science and Technology for Brain-inspired Intelligence, Fudan University, China\\
    $^6$School of Computer Science and Engineering, Southeast University, China \\
{\tt\small\{cziyuanyang,cyy262511,zwwang1228\}@gmail.com},\quad
{\tt\small chenyang.list@seu.edu.cn}\\
{\tt\small hmshan@fudan.edu.cn},\quad {\tt\small yzhang@scu.edu.cn}\\
    }

\begin{document}
\maketitle

{\blfootnote{
    $^*$Corresponding author.
    }}
    
\input{sec/0_abstract}    
\input{sec/1_intro}

\input{sec/2_formatting}
{
    \small
    \bibliographystyle{ieeenat_fullname}
    \bibliography{main}
}

\input{sec/X_suppl}

\end{document}

%% file: sec/0_abstract.tex
\begin{abstract}
Reducing radiation doses benefits patients, however, the resultant low-dose computed tomography (LDCT) images often suffer from clinically unacceptable noise and artifacts.
While deep learning~(DL) shows promise in LDCT reconstruction, it requires large-scale data collection from multiple clients, raising privacy concerns. Federated learning~(FL) has been introduced to address these privacy concerns; however, current methods are typically tailored to specific scanning protocols, which limits their generalizability and makes them less effective for unseen protocols.
To address these issues, we propose \textbf{SCAN-PhysFed}, a novel \textbf{SC}anning- and \textbf{AN}atomy-level personalized \textbf{Phy}sics-Driven \textbf{Fed}erated learning paradigm for LDCT reconstruction. 
Since the noise distribution in LDCT data is closely tied to scanning protocols and anatomical structures being scanned, we design a dual-level physics-informed way to address these challenges.
Specifically, we incorporate physical and anatomical prompts into our physics-informed hypernetworks to capture scanning- and anatomy-specific information, enabling dual-level physics-driven personalization of imaging features. These prompts are derived from the scanning protocol and the radiology report generated by a medical large language model (MLLM), respectively.
Subsequently, client-specific decoders project these dual-level personalized imaging features back into the image domain.
Besides, to tackle the challenge of unseen data, we introduce a novel protocol vector-quantization strategy (PVQS), which ensures consistent performance across new clients by quantifying the unseen scanning code as one of the codes in the scanning codebook.
Extensive experimental results demonstrate the superior performance of SCAN-PhysFed on public datasets\footnote{Our code has been released at \url{https://github.com/Zi-YuanYang/SCAN-PhysFed}.}.
﻿

\end{abstract}

%% file: sec/1_intro.tex
\section{Introduction}
\label{sec:intro}
Computed tomography (CT) is a key diagnostic tool as it non-invasively visualizes anatomical structures within the human body. Despite its clinical benefits, CT scans raise radiation concerns, which can lead to genetic, cancerous, and other health diseases.
Lowering CT radiation dose, through lowering incident photons or decreasing sampling views, has emerged as a promising solution to reduce radiation risks and accelerate scanning. However, reconstructed images under these conditions suffer from significant quality degradation, compromising clinical applicability~\cite{wang2008outlook}.

\begin{figure}[t]
  \centering
   \includegraphics[width=\linewidth]{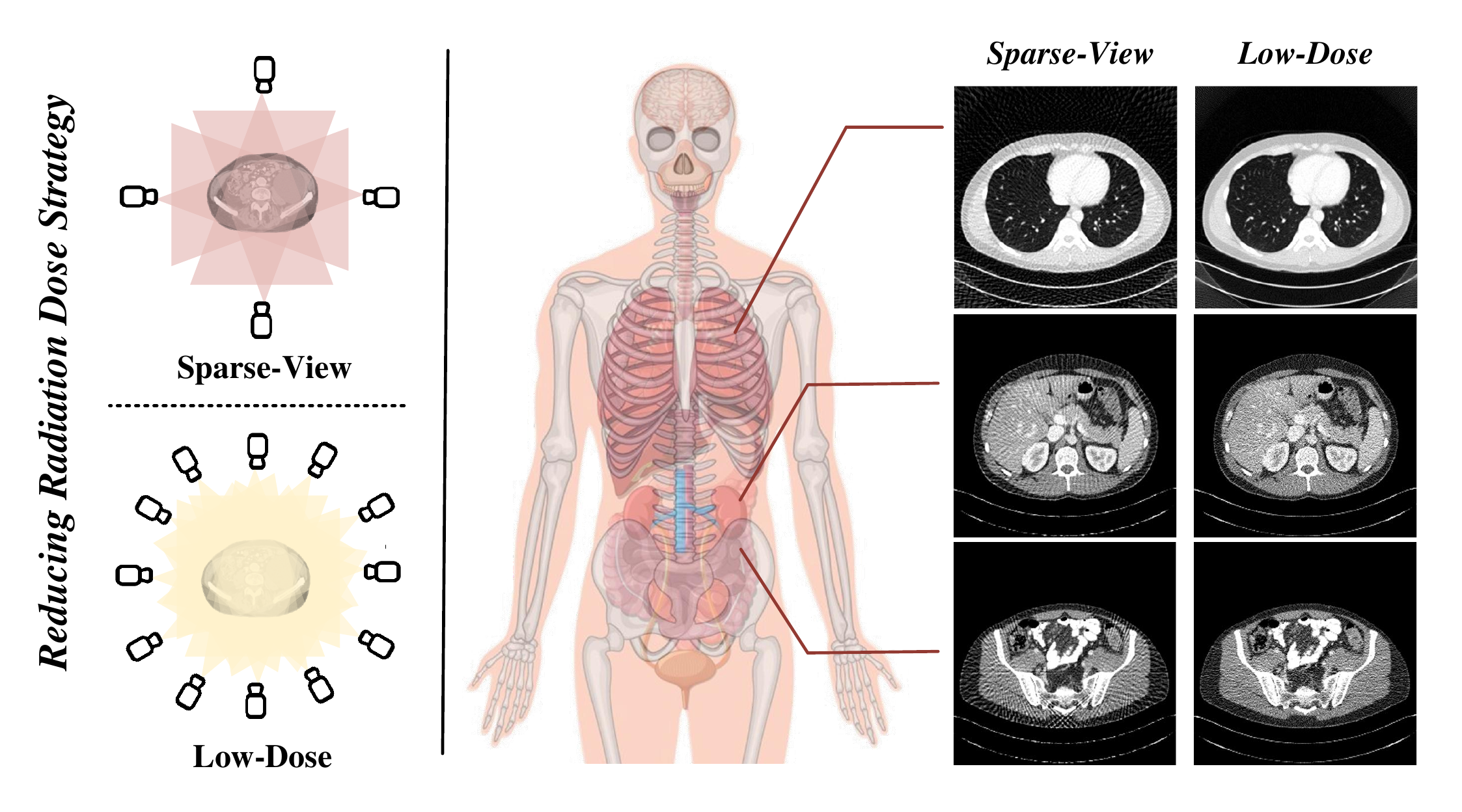}
   \vspace{-20pt}
   \caption{Illustration of different  radiation reduction strategies. ``Sparse-View'' and ``Low-Dose" represent  the scanning with sparse views and low incident photons, respectively.}
   \label{fig:toy}
   \vspace{-10pt}
\end{figure}

Recently, deep learning~(DL)-based methods have shown great promise for low-dose CT~(LDCT) imaging~\cite{wang2020deep}. However, most DL-based LDCT reconstruction methods are condition-specific and do not account for privacy concerns. Recently, researchers have attempted to introduce federated learning~(FL), a privacy-preserving distributed learning paradigm, in LDCT reconstruction to alleviate privacy concerns. 
Although these methods aim to enhance generalization, they still face significant challenges arising from data heterogeneity, caused by diverse noise distributions in CT imaging across different protocols and anatomical structures.
We believe that the further advancement of FL in LDCT reconstruction urgently requires a new framework capable of leveraging comprehensive physical information from the scanning process, thereby breaking through the current bottlenecks to achieve better results.


To tackle this issue, we aim to leverage the physical principles of the imaging process. During the projection process, the measurement is determined by the anatomical structures the X-Ray passed through, such as fat, bone, and fluid, each with a different attenuation coefficient.
Additionally, the system matrix is significantly affected by scanning protocols, including tube current, voltage settings, and scanning angles. These factors collectively impact the quality and characteristics of the reconstructed CT images during the backprojection phase, resulting in varying levels of noise and artifacts. An example is provided in Figure~\ref{fig:toy} to illustrate this more clearly.

By leveraging the physics-specific knowledge discussed earlier, we reimagine LDCT reconstruction within the federated learning framework. Our approach adopts a dual-level strategy to alleviate heterogeneity, addressing both scanning- and patient-level variations. In practice, scanning protocols, which serve as hyperparameters for scanners, are readily available, but anatomical information has traditionally been difficult to generate due to patient variability and the complexity of accurately modeling different tissue types. Recent advances in large language models (LLMs) offer new possibilities for automatically generating anatomical information, potentially improving the reconstruction.

Building on this, we propose \textbf{SCAN-PhysFed}, a novel \textbf{SC}anning- and \textbf{AN}atomy-level personalized \textbf{Phy}sics-Driven \textbf{Fed}erated learning paradigm for LDCT reconstruction. 
Specifically, we address the data heterogeneous problem in a dual-level strategy. 
A detailed radiology report is generated using a pretrained LLM. Then, a patient-level anatomy-informed hypernetwork is designed to create a modulation map from the radiology report that seamlessly integrates with the imaging features, aligning them with each patient’s unique anatomical structures.
Simultaneously, scanning features are integrated via a scanning-informed hypernetwork to personalize imaging features based on the protocol. To ensure that the scanning feature vectors are distinguishable, we apply an orthogonal loss during training to reinforce personalization. This dual-modulation strategy enables personalized CT imaging at both the patient and scanning levels, guided by the principles of physical imaging.



Additionally, existing personalized FL CT imaging works do not address the challenge of the unseen clients with varying scanning protocols. In this paper, we propose a \textbf{P}rotocol \textbf{V}ector \textbf{Q}uantization \textbf{S}trategy (\textbf{PVQS}) to quantize the unseen scanning protocol by matching it with the closest existing vectors, selecting the corresponding client-specific decoder. The main contributions of this paper are summarized as follows:
\begin{itemize}



\item We introduce LLM into LDCT reconstruction within the FL framework, proposing SCAN-PhysFed. To the best of our knowledge, this is the first work to apply LLM in this field.

\item To address the challenge of data heterogeneity, we draw inspiration from the physical process and propose a dual-level solution, achieving personalized CT imaging at both the scanning and patient levels.

\item We introduce a novel strategy, PVQS, which ensures our method remains effective for unseen clients with varying scanning protocols, a critical challenge often overlooked by previous works.
    
\end{itemize}

\section{Related Works}
\textbf{LDCT Imaging.} Traditional reconstruction methods, primarily rely on the sparsity of the image~\cite{ravishankar2019image}. However, these methods are generally time-consuming. Recently, deep learning-based methods have shown faster and more accessible performance, gaining significant attention in this field. For example, Chen~\etal~\cite{chen2017low} introduced residual structure into the autoencoder and proposed encoder-decoder convolutional neural network (RED-CNN) for LDCT denoising. You~\etal~\cite{you2019ct} extended RED-CNN by incorporating a generative adversarial network (GAN) for super resolution. Additionally, researchers also explored the use of transformer blocks to achieve robust performance, such as in Uformer~\cite{wang2022uformer}, TransCT~\cite{zhang2021transct}, and RegFormer~\cite{xia2023regformer}. Recently, some researchers have proposed diffusion-based CT reconstruction methods~\cite{xia2022low, gao2023corediff,lu2024pridediff}, which show promising results but require substantial computational cost.
Moreover, the methods mentioned above are tailored to specific scanning protocols and depend on data collection from multiple clients, raising significant privacy concerns.

\begin{figure*}[t]
  \centering
   \includegraphics[width=\linewidth]{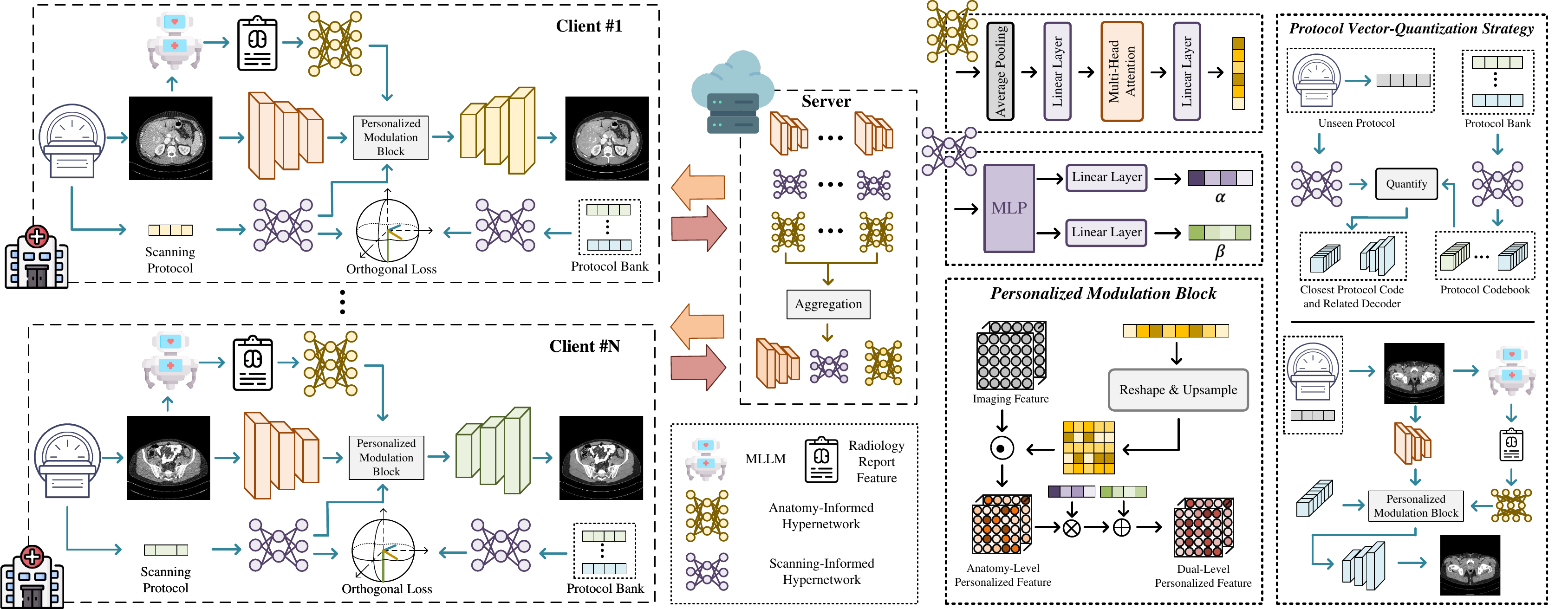}
   \vspace{-15pt}
   \caption{The overall learning paradigm of our proposed SCAN-PhysFed.}
   \label{fig:sample}
   \vspace{-15pt}
\end{figure*}

\noindent \textbf{Federated Learning.}
In recent years, FL has gained increasing attention as it offers a decentralized approach to DL while preserving data privacy~\cite{ye2023heterogeneous}. FedAvg~\cite{mcmahan2017communication} is the earliest FL method, which proposed a server-client framework to learn a global-shared model by averaging the local models of different parties. However, this method suffers from the model shift problem caused by serious statistic heterogeneity~\cite{karimireddy2020scaffold}. To alleviate this concern, recent works have proposed methods to obtain a more robust global-shared model, such as FedProx~\cite{li2020federated}, MOON~\cite{li2021model}, and FedNova~\cite{wang2020tackling}. For serious non-iid data distributions, a single shared global model is insufficient to address the variability across different datasets. As a result, several personalized FL methods have been proposed~\cite{li2021fedbn, arivazhagan2019federated, yang2024fedas, deng2024fedasa}.

FL has recently been introduced in healthcare due to its privacy-preserving nature~\cite{wu2022communication, yang2023dynamic, guan2024federated}. Several studies have focused on medical imaging~\cite{feng2022specificity, li2023semi}. For example, Chen~\etal~\cite{chen2023federated} proposed a cross-domain learning approach, including individual-client sinogram learning and cross-client image reconstruction learning. Yang~\etal~\cite{yang2023hypernetwork} proposed to use a client-specific hypernetwork to modulate the global-shared data. Chen~\etal~\cite{chen2024fedfdd} proposed utilizing separate networks to address the high- and low-frequency components of CT data. While these methods have achieved promising performance, they still face significant challenges from data heterogeneity and overlook the issue of unseen protocols.

\noindent \textbf{Large Language Model.}
The emergence of large language models~(LLMs), such as GPT~\cite{achiam2023gpt} and Llama~\cite{touvron2023llama}, has marked a new era in artificial intelligence. Recently, large-scale medical LLMs~(MLLMs) have been introduced specifically for healthcare applications~\cite{liu2023medical, liu2024moe, zhu2024llafs}. These models demonstrate significant potential in understanding and reasoning over medical data, excelling in tasks such as visual question answering (VQA) and text generation~\cite{moor2023foundation}. For instance, Alkhaldi \etal~\cite{alkhaldi2024minigpt} developed Minigpt-med, which is capable of handling multiple tasks, including VQA, medical report generation, and disease identification from medical imagery. Despite MLLMs being used in many other downstream tasks~\cite{kumar2024medvisionllama}, there has been no attempt to leverage them to address the domain shift problem in CT reconstruction within FL settings.

%% file: sec/2_formatting.tex
\section{Methodology}
\label{sec:metho}


\subsection{Overview}
This paper aims to address generalization challenges in FL for LDCT reconstruction, focusing on alleviating data heterogeneity within existing clients and maintaining performance for new clients.
Figure~\ref{fig:sample} illustrates the overall learning paradigm of the proposed SCAN-PhysFed framework. Specifically, robust imaging representations are obtained using a collaboratively learned shared encoder across clients. To tackle the heterogeneity issue, two physics-informed hypernetworks are designed to capture scanning- and anatomy-specific information, personalizing the imaging features through a specialized modulation block.
Subsequently, client-specific decoders project the personalized imaging features back to the image domain. For the $i$-th client, the reconstruction process is formulated as:
\begin{equation}
\mat{\hat{Y}} = D_i(P(f_\mat{X},H_s(g;\theta_{H_s}),H_a(f_t;\theta_{H_a}));\theta_{D_i}),
\label{eq:paradigm}
\end{equation}
where $f_\mat{X}$ denotes the imaging feature of $\mat{X}$ from the shared encoder $E$, and $\mat{\hat{Y}}$ is the corresponding normal-dose CT image. $H_s$ and $H_a$ are the scanning- and anatomy-informed hypernetworks, parameterized by $\theta_s$ and $\theta_a$, respectively. $g$ and $f_t$ represent the scanning protocol vector and the radiology report feature derived from the MLLM. $P$ denotes the personalized modulation block, and $D_i$ denotes $i$-th client decoder parametrized by $\theta_{D_i}$.

Additionally, to address the challenge of unseen data, we introduce a novel protocol vector-quantization strategy~(PVQS), which ensures consistent performance across previously unseen clients by quantizing the unseen scanning code as one of the codes in the scanning codebook.




\subsection{Personalized Modulation Block}

\noindent\textbf{Anatomy-Level Personalization.} 
The patient’s unique anatomy significantly affects noise distribution during the imaging process, making it an essential factor to consider.
To address this issue, we leverage MLLM, which excels at training on large-scale datasets and demonstrates strong performance in radiology report generation. Specifically, we adapt miniGPT-Med~\cite{alkhaldi2024minigpt} to generate radiology reports from CT images, using the text encoder’s feature output $f_t\in \mathbb{R}^{1\times d}$ as a prompt for our anatomy-informed hypernetwork, where $d$ denotes the feature dimension.

Since the dimensionality of $f_t$ is high, we first apply a pooling operation to reduce it. The process is formulated as follows:
\begin{equation}
    f_{tl} = \operatorname{Linear}(\operatorname{AvgPool}(f_t)),
\end{equation}
where $\mathrm{AvgPool}$ denotes an average pooling operation that reduces the feature size by a factor of $1/4$. A linear layer is then applied to map the feature to $f_{tl}\in \mathbb{R}^{1\times h}$, where $h$ denotes the hidden dimension.

Afterward, we establish long-range dependencies within $f_{tl}$ using the multi-head attention mechanism, followed by a linear layer to extract the anatomy feature $f_{an}$.
This process can be formulated as:
\begin{equation}
    f_{an} = \operatorname{Linear}(\operatorname{Multi-Head}(f_{tl})).
\end{equation}


After this, $f_{an}$ encapsulates anatomy-specific information tailored to different CT data. Since this feature reflects anatomy-specific details, we use it to spatially personalize the imaging feature $f_\mat{X}$. The process is formulated as:
\begin{equation}
    f_{ana} = f_\mat{X} \odot \operatorname{Reshape}(f_{an}),
\end{equation}
where $f_\mat{X}$ denotes the imaging feature, $\operatorname{Reshape}$ represents the reshape operation to match the dimension, and ``$\odot$" denotes the Hadamard product. $f_{ana}$ is the anatomy-level personalized feature.




\noindent\textbf{Scanning-Level Personalization.}
In addition to anatomical variations affecting noise distribution, the physical imaging process shows that noise distribution is largely determined by the scanning protocol, which involves several key physical parameters in CT scanners, including the number of views, the number of detector bins, pixel length, detector bin length, the distance between the source and rotation center, the distance between the detector and rotation center, and the photon number of incident X-rays. 

Then, we utilize $g$, a vector containing these parameters, to construct our scanning-level prompt, capturing the latent relationship between the scanning protocol and noise distribution for the imaging network. A detailed explanation of $g$ can be found in Appendix B.

Given the low dimensionality of the physical parameters, we employ a Multi-Layer Perceptron (MLP) with three linear layers to capture and represent the underlying physical information. Two modulation parameter vectors are then extracted to capture scanning-level personalization knowledge. This process can be formulated as follows:
\begin{equation}
    \alpha, \beta = H_s(g; \theta_s),
\end{equation}
where $\alpha$ and $\beta$ are the modulation parameter vectors, which are used to inject the scanning-level information to personalize the imaging feature as:
\begin{equation}
    f_{per} = \alpha \otimes f_{ana} + \beta,
\end{equation}
where $f_{per}$ is the dual-level personalized imaging feature and $\otimes$ denotes the multiplication operation along the channel dimension. This approach modulates the imaging feature using the scanning prior.

Our proposed personalization block imposes no requirements on the imaging network architecture, so the dimensions of personalization parameters should be fine-tuned accordingly.

\subsection{Protocol Vector-Quantization Strategy}
Existing FL CT imaging methods have not addressed the challenge of unseen data, resulting in degraded performance in new domains. To overcome this limitation and ensure robust performance for unseen protocols, we introduce PVQS. While the patient anatomy remains largely consistent, the primary noise variation between unseen and existing clients is driven by the low-dose strategy. Therefore, PVQS is designed to enable SCAN-PhysFed to maintain performance across diverse, unseen low-dose protocols.

Specifically, in PVQS, the protocol codes for different existing scanning protocols, generated by the MLP in $H_s$, are stored in a protocol codebook. For an unseen protocol, we first quantize it as one of the code vectors in the protocol codebook. The process can be calculated as follows:
\begin{equation}
    i^{*}=\arg \min _{i}\left(1-\frac{\vct{c}_{un} \cdot \vct{c}_{i}}{\|\vct{c}_{un}\|\left\|\vct{c}_{i}\right\|}\right),
\end{equation}
where $\vct{c}_{un}$ and $\vct{c}_{i}$ denote the protocol code of the unseen protocol and the $i$-th client's protocol, respectively. $i^*$ indicates the index of the closest code.

The unseen protocol is not used as the input to $H_s$; instead, we directly use $\vct{c}_{i^*}$ to replace the unseen protocol code to personalize the imaging feature and use the corresponding decoder $D_{i^*}$ to project the personalized features back to the image domain.

In this way, PVQS can effectively avoid potential distribution shifts in unseen data and maintain a consistent feature space across domains, ensuring reliable and robust performance for unseen protocols.

\begin{table*}[htbp]
\centering
\caption{The Quantitative Results (PSNR (dB) and SSIM (\%)) for CNN-based LDCT Imaging Method.}
\label{tab:red}
\resizebox{\textwidth}{!}{%
\begin{tabular}{@{}lllllllllllllllll|ll@{}}
\hline
\centering
\multirow{2}{*}{} & \multicolumn{2}{c}{Client \#1} & \multicolumn{2}{c}{Client \#2} & \multicolumn{2}{c}{Client \#3} & \multicolumn{2}{c}{Client \#4} & \multicolumn{2}{c}{Client \#5}  & \multicolumn{2}{c}{Client \#6} & \multicolumn{2}{c}{Client \#7} & \multicolumn{2}{c}{Client \#8} & \multicolumn{2}{|c}{Average} \\
                  & PSNR         & SSIM          & PSNR        & SSIM         & PSNR         & SSIM         & PSNR         & SSIM         & PSNR         & SSIM        & PSNR         & SSIM        & PSNR         & SSIM    & PSNR & SSIM & PSNR & SSIM      \\ \hline
\multicolumn{19}{c}{Generic FL Methods} \\
\hline
                  FedAvg~\cite{mcmahan2017communication}& 29.99 & 82.20 & 32.82 & 78.23 & 34.68 & 86.99 & 32.30 & 80.63 & 36.87 & 92.06 & 37.36 & 89.02 & 32.73 & 80.66 & 35.46 & 87.65 & 34.03 & 84.70\\
                   \rowcolor{lightgray}
                  FedProx~\cite{li2020federated} & 29.94 & 83.12 & 32.69 & 75.46 & 36.95 & 91.58 & 32.32 & 80.91 & 38.95 & 95.34 & 37.58 & 88.13 & 32.94 & 81.62 & 35.86 & 88.65 & 34.65 & 85.60 \\
                  FedNova~\cite{wang2020tackling} & 29.95 & 82.29 & 32.57 & 80.57 & 37.77 & 92.16 & 32.09 & 78.98 & 34.42 & 88.69 & 36.05 & 89.57 & 32.81 & 81.41 & 35.89 & 88.71 & 33.94 & 85.30  \\
                   \rowcolor{lightgray}
                  MOON~\cite{li2021model} & 32.50	& 75.61 & 35.55 & 85.41 & 35.72 & 80.30 & 35.69 & 82.10 & 35.38 & 78.47 & 36.31 & 84.35 & 35.70 & 81.17 & 35.62 & 80.40 & 35.31 & 80.98 \\
                  FedDG~\cite{liu2021feddg} & 33.37 & 81.40 & 34.01 & 82.39 & 33.56 & 76.37 & 34.43 & 81.49 & 34.06 & 76.70 & 33.97 & 80.17 & 34.75 & 80.87 & 34.25 & 79.23 & 34.05 & 79.83 \\
                   \rowcolor{lightgray}
                  FedKD~\cite{wu2022communication}                  &  29.74 & 83.11 & 32.17 & 71.65 & 36.65 & 90.93 & 32.10 & 79.19 & 40.58 & 96.71 & 37.07 & 82.55 & 32.88 &	81.30 & 36.28 & 89.08 & 34.68 & 84.31  \\
                  \hline 
\multicolumn{19}{c}{Personalized FL Methods} \\
\hline 
                  FedPer~\cite{arivazhagan2019federated}  & 33.23	& 88.20 & 35.06 & 84.89 & 38.79 & 92.70 & 35.58 & 86.12 & 41.76 &	96.97 & 36.85 & 89.96 & 35.69 & 86.30 & 38.48 & 91.69 & 36.93 & 89.60      \\
                   \rowcolor{lightgray}
                  FedBN~\cite{li2021fedbn} & 32.55 & 84.59 & 34.76 & 84.15 & 38.18 & 90.30 & 34.77 & 83.30 & 41.24 & 96.11 & 38.66 & 91.19 & 34.74 & 83.35 & 37.89 & 90.35 & 36.60 & 87.92  \\
                  FedMRI~\cite{feng2022specificity} & 33.28 & 83.43 & 34.51 & 85.81 & 38.62 & 92.47 & 35.86 & 87.21 & 41.49 & 96.27 & 38.22 & 91.32 & 35.87 & 86.34 & 38.09 & 91.94 & 36.99 & 89.35 \\
 \rowcolor{lightgray}
                  HyperFed~\cite{yang2023hypernetwork} & 33.74 & 88.57 & 34.83 & 85.05 & 38.87 & 92.78 & 35.44 & 85.40 & 42.94 & 97.48 & 39.15 & 92.96 & 34.99 & 85.19 & 38.12 & 91.29 & 37.26 & 89.84
                  \\
                  FedFDD~\cite{chen2024fedfdd} & \textbf{36.18} & 93.67 & 38.43 & 94.19 & 39.72 & 96.47 & 38.51 & 95.30 & 45.56 & 98.73 & 42.72 & 97.61 & 40.74 & 96.01 & 38.54 & 94.39 & 40.05 & 95.79\\
                  \hline 
 \rowcolor{lightgray}
                  SCAN-PhysFed & 34.49 & \textbf{96.40} & \textbf{39.21} & \textbf{96.09} & \textbf{40.45} & \textbf{97.93} & \textbf{40.51} & \textbf{97.61} & \textbf{47.81} & \textbf{99.14} & \textbf{44.08} & \textbf{98.51} & \textbf{42.54} & \textbf{97.86} & \textbf{40.02} & \textbf{97.37} & \textbf{41.14} & \textbf{97.62} \\               
\hline 
\end{tabular}
}
\vspace{-10pt}
\end{table*}

\subsection{Learning Paradigm}

Since the noise distribution varies across protocols, it is challenging to use a single shared imaging network to accommodate all variations. In the context of FL for LDCT imaging, as assumed in previous works~\cite{yang2023hypernetwork,chen2024fedfdd}, the optimization challenge consists of two main components: imaging feature extraction and personalized projection. Specifically, for feature extraction, we propose aggregating client-side encoders to capture robust, comprehensive imaging features. To fully leverage diverse data across clients, we globally share the hypernetworks in SCAN-PhysFed. However, projecting imaging features from different domains into a consistent clean representation is challenging with a single shared decoder. Therefore, we propose protocol-specific decoders to effectively handle variations in scanning protocols across clients.

In PVQS, the unseen protocol code is quantized using the codebook. Therefore, it is essential to ensure the protocol code is discriminative and supports comprehensive personalization information. To achieve this, we introduce an orthogonal loss $\mathcal{L}_{\text{orth}}$ for the $i$-th client, defined as follows:
\begin{equation}
    \mathcal{L}_{\text{orth}}=\sum_{j=1}^{K}\left|\vct{c}_{i} \cdot \vct{c}_{j}\right|^{2} \quad \textit{s.t.}\quad i\neq j,
\end{equation}
where $K$ is the number of clients (protocols).

The Mean Squared Error (MSE) is used as the imaging loss, and the total loss is formulated as:
\begin{equation}
    \mathcal{L}_{\text{total}} = \mathcal{L}_{\text{MSE}}(\mat{Y},\hat{\mat{Y}}) + \tau \mathcal{L}_{\text{orth}},
\end{equation}
where $\mat{Y}$ is the reconstructed image, and $\tau$ is the temperature weight of $\mathcal{L}_{\text{orth}}$.

To help readers understand the implementation details, we also provide the pseudocode of our method in Appendix~A.






\begin{figure}[t]
  \centering
   \includegraphics[width=\linewidth]{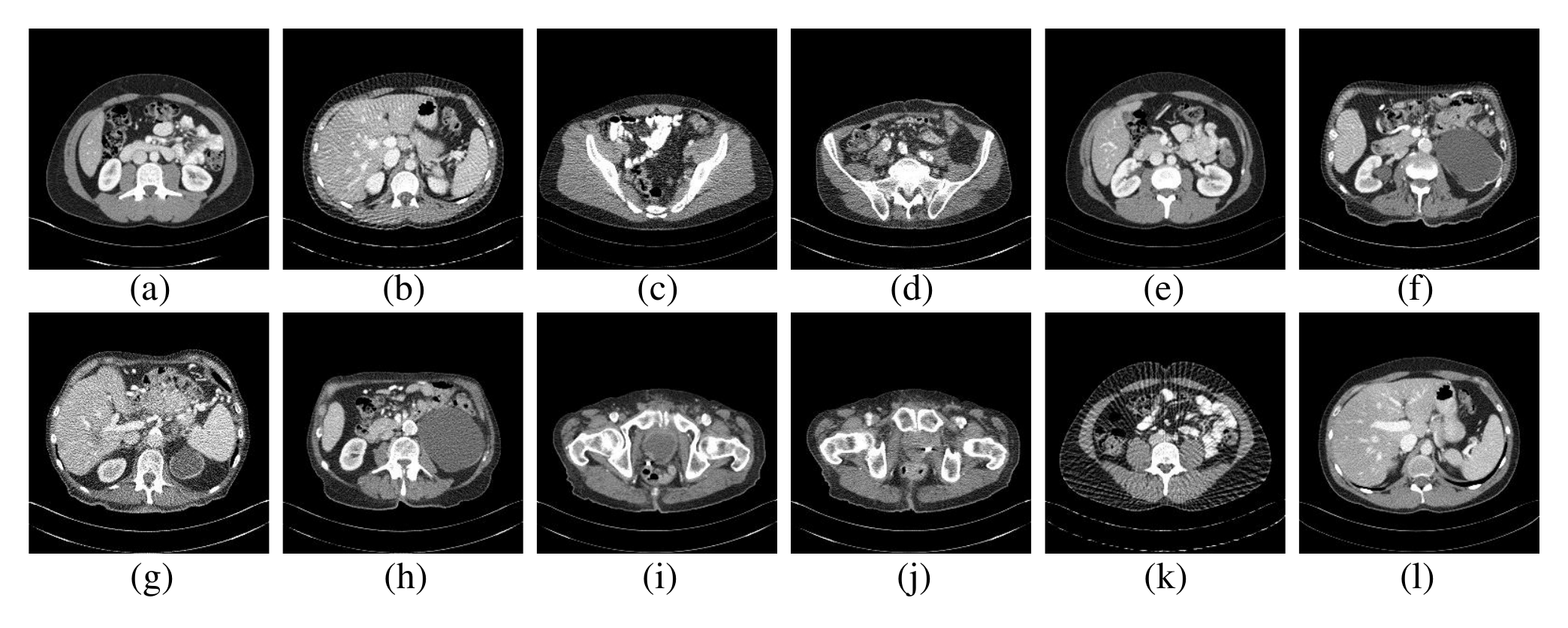}
   \vspace{-15pt}
   \caption{Simulated examples under different protocols. (a)-(h) show examples from training clients with known protocols, while (i)-(l) present examples from unseen protocols.}
   \label{fig:example}
   \vspace{-10pt}
\end{figure}

\begin{figure*}[t]
  \centering
   \includegraphics[width=\linewidth]{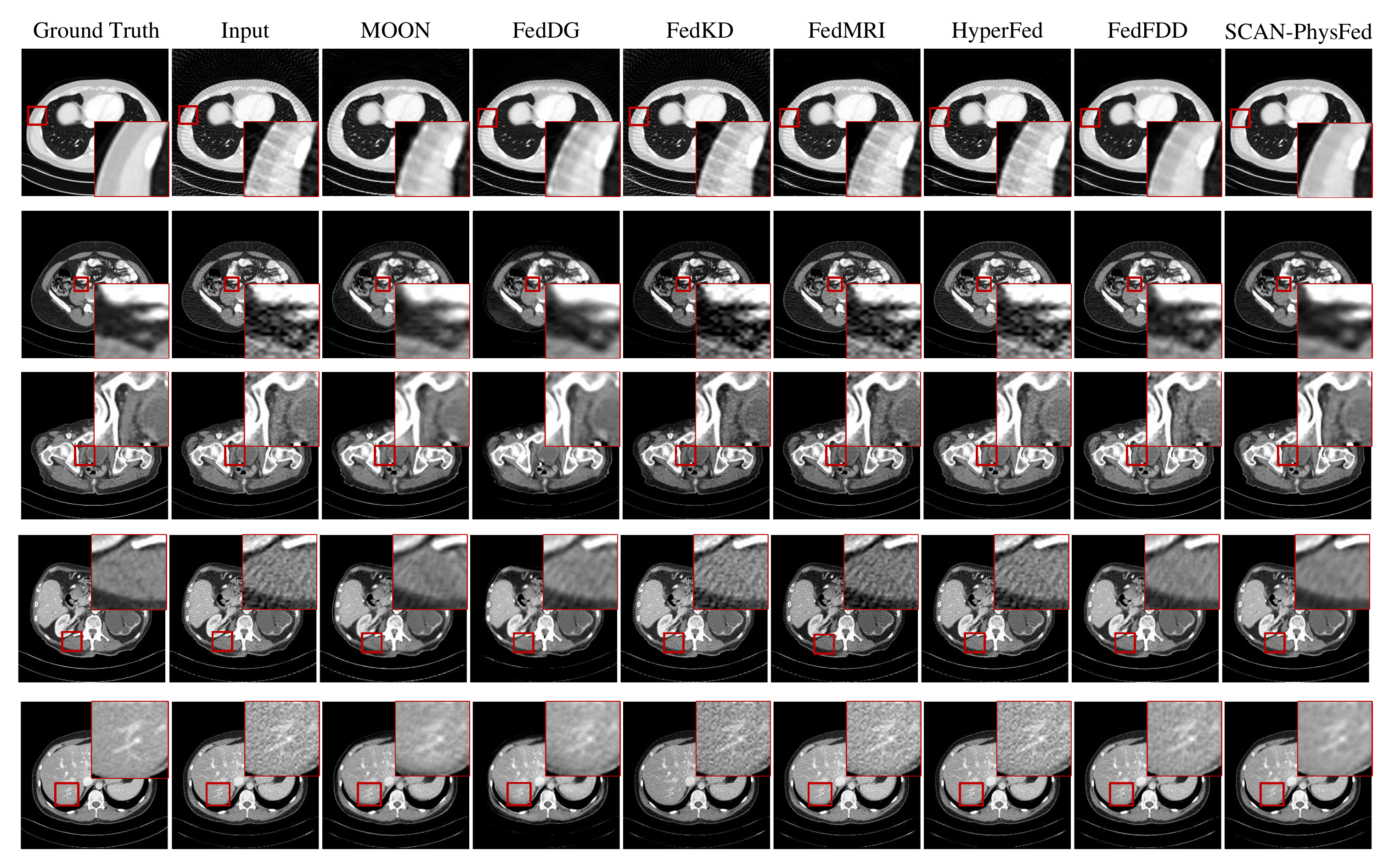}
   \vspace{-15pt}
   \caption{Qualitative results of six selected comparison methods and our method across different clients using the classical convolutional-based LDCT imaging network. Rows one to five represent Clients \#2, \#3, \#5, \#6, and \#7, respectively. The display window for the first row is [-1024, 200] HU, while for the other rows, it is [-160, 240] HU.}
   \label{fig:red}
   \vspace{-10pt}
\end{figure*}

\begin{figure}[htbp]
  \centering
   \includegraphics[width=\linewidth]{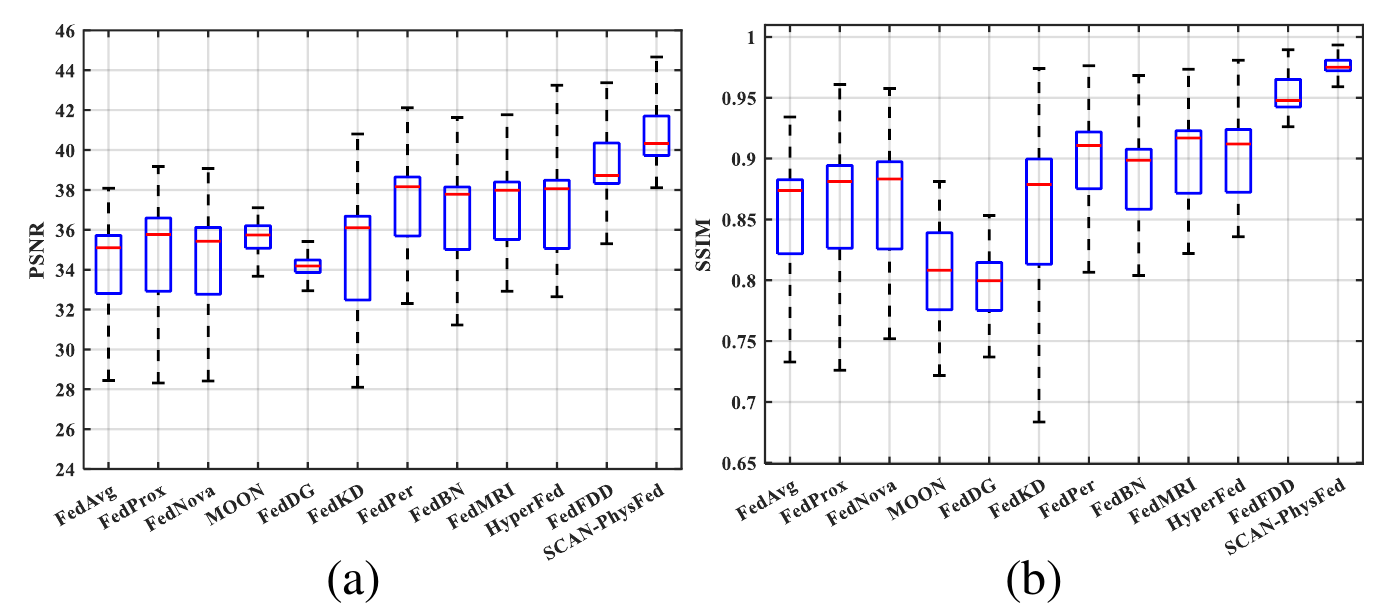}
   \vspace{-15pt}
   \caption{Boxplots of the average results across all clients. (a) and (b) represent PSNR and SSIM, respectively.}
   \label{fig:boxplot}
   \vspace{-15pt}
\end{figure}

\section{Experiments}
\textbf{Experimental Setting.}
The proposed method was implemented using PyTorch and optimized by Adam~\cite{kingma2014adam} at a learning rate of 0.001. The number of communication rounds was set to 200 and the batch size was set to 20. The experiments were conducted in an environment with an AMD Ryzen 7 5800X CPU and one single NVIDIA GTX 3080 Ti GPU.

\noindent \textbf{Dataset.} The “2016 NIH-AAPM-Mayo Clinic Low-Dose CT Grand Challenge” dataset \cite{mccollough2016tu}, which contains 5,936 full-dose CT images from 10 patients, was used to evaluate the proposed method. In this paper, 8 patients were used for training and 2 for testing. To simulate a realistic scenario, we ensure that the patients for training do not overlap across different clients. The two patients for testing were simulated using different protocols to form the testing set.

Simulated examples are shown in Figure~\ref{fig:example}, where noticeable noise heterogeneity among clients can be observed, caused by differences in anatomy and protocols. The detailed dataset preparation steps and scanning protocols used are provided in Appendix B.

\noindent \textbf{Baselines.}
We compared our SCAN-PhysFed approach with both generic and personalized FL methods. For generic FL methods, we include FedAvg~\cite{mcmahan2017communication}, FedProx~\cite{li2020federated}, FedNova~\cite{wang2020tackling}, MOON~\cite{li2021model}, FedDG~\cite{liu2021feddg}, and FedKD~\cite{wu2022communication}. For personalized FL methods, we evaluated FedPer~\cite{arivazhagan2019federated}, FedBN~\cite{li2021fedbn}, FedMRI~\cite{feng2022specificity}, HyperFed~\cite{yang2023hypernetwork}, and FedFDD~\cite{chen2024fedfdd}. To ensure fairness, we kept the training settings consistent across all methods.

\noindent \textbf{Evaluation Metrics.} We use peak signal-to-noise ratio~(PSNR) to evaluate pixel-wise accuracy and structural similarity~(SSIM) to assess perceived visual quality~\cite{wang2004image}. For both metrics, higher values indicate better performance.

\subsection{Comparison with other methods}

We evaluate our method alongside other FL methods. To ensure fairness, we use the same imaging network, RED-CNN~\cite{chen2017low}, which is one of the most powerful models in CT imaging. Quantitative results are provided in Table~\ref{tab:red}. Our method achieves the best performance across most clients, outperforming other generic and personalized FL methods in terms of average PSNR and SSIM.
Furthermore, unlike other personalized methods that show instability across clients, such as HyperFed in Client \#2 / \#7 and FedFDD in Client \#4, our method maintains consistent performance across varying protocols. We also observe that sparse-view protocols (Clients \#2, \#4, and \#6) present challenges for both generic and personalized FL methods, with significant performance gaps in these protocols compared to others. However, by incorporating dual-level physical prior knowledge, our method maintains robust performance.


\begin{table*}[htbp]
\centering
\caption{Quantitative Results of PSNR and SSIM for Transformer-based LDCT Imaging Method.}
\label{tb:uformer}
\resizebox{\textwidth}{!}{%
\begin{tabular}{@{}lllllllllllllllll|ll@{}}
\hline
\centering
\multirow{2}{*}{} & \multicolumn{2}{c}{Client \#1} & \multicolumn{2}{c}{Client \#2} & \multicolumn{2}{c}{Client \#3} & \multicolumn{2}{c}{Client \#4} & \multicolumn{2}{c}{Client \#5}  & \multicolumn{2}{c}{Client \#6} & \multicolumn{2}{c}{Client \#7} & \multicolumn{2}{c}{Client \#8} & \multicolumn{2}{|c}{Average} \\
                  & PSNR         & SSIM          & PSNR        & SSIM         & PSNR         & SSIM         & PSNR         & SSIM         & PSNR         & SSIM        & PSNR         & SSIM        & PSNR         & SSIM    & PSNR & SSIM & PSNR & SSIM      \\ \hline
\multicolumn{19}{c}{Generic FL Methods} \\
\hline
                  FedAvg~\cite{mcmahan2017communication} & 34.31 & 87.88 & 38.06 & 90.54 & 43.51 & 97.56 & 41.05 & 95.43 & 45.24 & 98.47 & 43.15 & 97.09 & 41.45 & 95.95 & 42.88 & 96.16 & 41.21 & 94.89\\        
                  \rowcolor{lightgray}
                  FedProx~\cite{li2020federated} & 34.32 & 87.01 & 37.74 & 88.89 & 43.73 & 97.95 & 41.12 & 93.92 & 45.39 & 98.61 & 42.99 & 97.08 & 41.38 & 94.31 & 42.10 & 94.55 & 41.10 & 94.04\\
  
                  FedNova~\cite{wang2020tackling} & 29.55 & 78.44 & 31.38 & 68.13 & 37.07 & 82.47 & 31.79 & 73.45 & 40.61 & 88.80 & 35.46 & 79.77 & 32.38 & 73.66 & 35.77 & 82.22 & 34.25 & 78.37 \\                  
                  \rowcolor{lightgray}
                  MOON~\cite{li2021model} & 34.28 & 93.20 & 38.09 & 91.17 & 43.64 & 97.77 & 41.09 & 95.66 & 45.25 & 98.43 & 42.82 & 96.91 & 41.42 & 95.97 & 42.20 & 96.18 & 41.10 & 95.66\\
                  
                  FedKD~\cite{wu2022communication} & 35.92 & 89.93 & 38.70 & 91.21 & 44.26 & 98.13 & 42.30 & 95.99 & 45.93 & 98.75 & 43.57 & 97.41 & 42.54 & 96.22 & 43.51 & 96.23 & 42.09 & 95.48 \\
                                    \hline
\multicolumn{19}{c}{Personalized FL Methods} \\
\hline
                  \rowcolor{lightgray}
                  FedPer~\cite{arivazhagan2019federated} & 35.37 & 93.18 & 39.03 & 93.55 & 42.75 & 97.13 & 40.88 & 96.20 & 46.55 & 98.83 & 42.87 & 96.68 & 41.87 & 96.70 & 40.79 & 94.78 & 41.26 & 95.82 \\

                  FedMRI~\cite{feng2022specificity} & 34.60 & 93.67 & 38.04 & 94.55 & 42.94 & 97.44 & 39.34 & 95.78 & 46.79 & 98.91 & 42.82 & 97.30 & 40.81 & 95.32 & 40.85 & 95.70 & 40.78 & 96.01 \\
                  \rowcolor{lightgray}
                  HyperFed~\cite{yang2023hypernetwork} & 34.60 & 87.59 & 38.62 & 91.08 & 44.13 & 97.61 & 42.25 & 96.24 & 45.92 & 98.59 & 43.71 & 97.37 & 42.44 & 96.66 & \textbf{43.56} & 96.63 & 41.90 & 95.22 \\

                  FedFDD~\cite{chen2024fedfdd} & 37.35 & \textbf{96.20} & 39.67 & 95.18 & 44.31 & 98.26 & 39.00 & 95.21 & 47.89 & 99.20 & 43.81 & \textbf{98.06} & 42.17 & \textbf{97.40} & 40.50 & 95.49 & 41.84 & 96.89 \\
                  \hline 
                  \rowcolor{lightgray}
                  SCAN-PhysFed  & \textbf{38.55} & 94.06 & \textbf{41.74} & \textbf{96.20} & \textbf{45.52} & \textbf{98.52} & \textbf{42.48} & \textbf{96.55} & \textbf{49.50} & \textbf{99.39} & \textbf{45.32} & 97.93 & \textbf{43.35} & 96.73 & 42.71 & \textbf{96.79} & \textbf{43.65} & \textbf{97.03}\\
\hline
\end{tabular}
}
\vspace{5pt}
\end{table*}

\begin{figure*}[t]
  \centering
   \includegraphics[width=\linewidth]{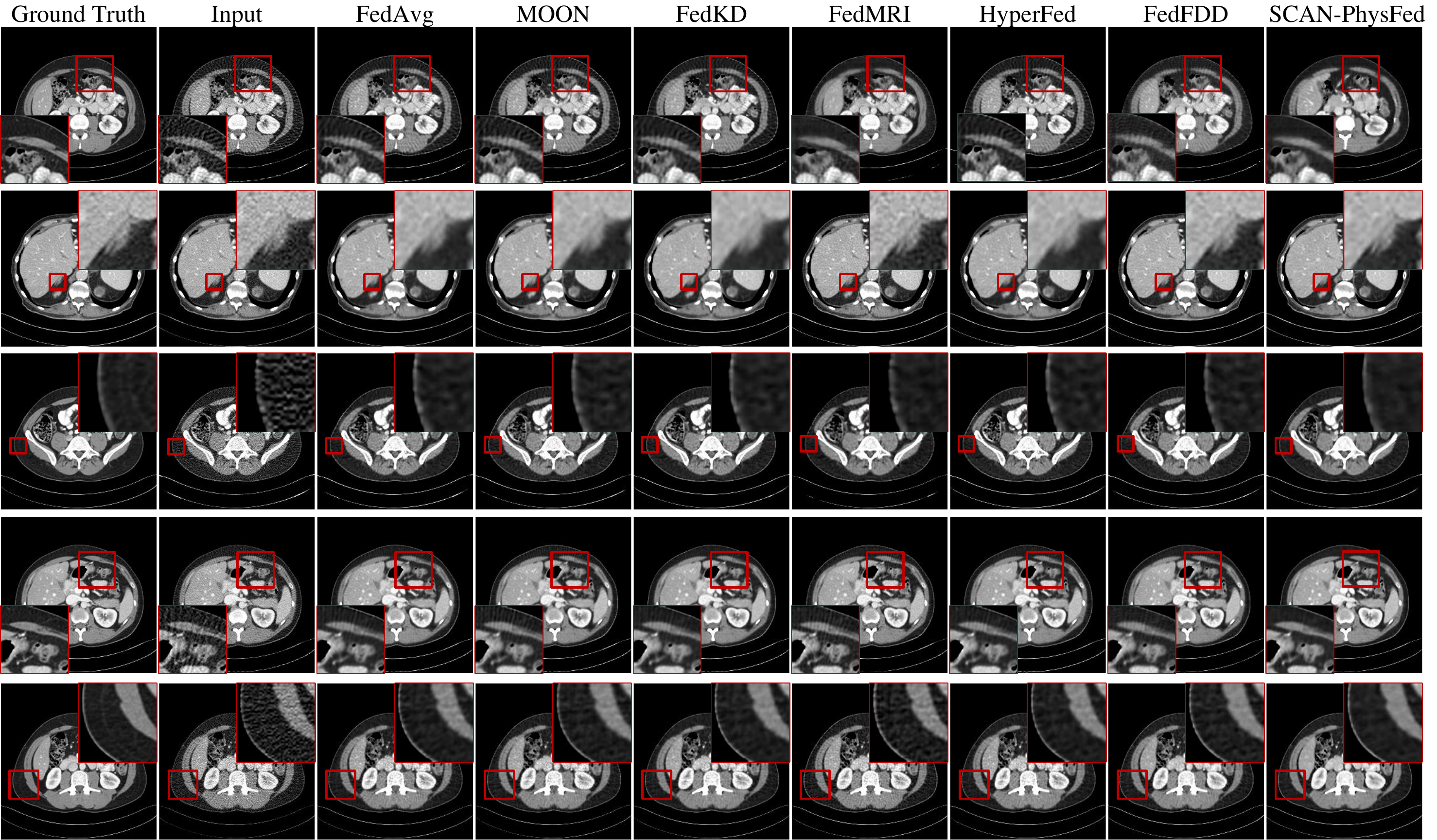}
   \caption{Qualitative results of the compared methods across different clients based on the transformer-based imaging network. Rows one to five represent Clients \#2, \#3, \#5, \#6, and \#7, respectively. The display window is set to [-160, 240] HU.}
   \label{fig:uformer_sample}
\end{figure*}

Representative slices produced by different methods are shown in Figure~\ref{fig:red}. We compared the top three methods from both generic and personalized FL approaches. Representative clients were selected based on different radiology reduction strategies: sparse-view (Clients \#2 and \#6) and low-dose (Clients \#3, \#5, and \#7). It can be observed that other methods exhibit noticeable noise or artifacts in sparse-view protocols, and fail to preserve details in low-dose protocols. However, ours effectively reduces noise and artifacts in the sparse-view protocols while preserving fine details in low-dose protocols. SCAN-PhysFed achieves strong performance across slices from different body parts, benefiting from the introduction of anatomy prior. Both quantitative and qualitative results demonstrate the effectiveness of incorporating physical information.

Additionally, we present boxplots of the results across all clients for different methods in Figure~\ref{fig:boxplot} to validate their stability. It can be seen that SCAN-PhysFed  outperforms all other methods in both PSNR and SSIM scores. In both plots, SCAN-PhysFed demonstrates a higher median and a narrower interquartile range, indicating strong stability in restoring LDCT data with varying noise distributions across different clients. Individual boxplots for each client are provided in Appendix C.

\begin{table*}[htbp]
\centering
\caption{The Quantitative Results of PSNR and SSIM for the Unseen Clients.}
\label{tb:open}
\tiny
\resizebox{.8\textwidth}{!}{%
\begin{tabular}{@{}lllllllll|ll@{}}
\hline
\centering
\multirow{2}{*}{} & \multicolumn{2}{c}{Unseen Client \#1} & \multicolumn{2}{c}{Unseen Client \#2} & \multicolumn{2}{c}{Unseen Client \#3} & \multicolumn{2}{c}{Unseen Client \#4} & \multicolumn{2}{|c}{Average}
\\
                  & PSNR         & SSIM          & PSNR        & SSIM         & PSNR         & SSIM         & PSNR         & SSIM         & PSNR         & SSIM        
                  \\ \hline
\multicolumn{11}{c}{Generic FL Methods} 
\\
\hline
  FedAvg~\cite{mcmahan2017communication}  & 35.57 & 88.68 & 34.00 & 85.26 & 32.52 & 76.71 & 36.86 & 91.44 & 32.89 &81.35 \\
                    \rowcolor{lightgray}
                  FedProx~\cite{li2020federated} & 
                  36.69 & 91.16 & 34.30 & 86.39 & 32.58 & 72.13 & 39.13 & 94.87 & 34.74 & 85.52  \\
                  FedNova~\cite{wang2020tackling}  & 35.45 & 88.83 & 34.23 & 86.29 & 32.39 & 69.20 & 35.40 & 89.99 & 34.38 & 83.58 \\
                                    \rowcolor{lightgray}
                  MOON~\cite{li2021model} & 35.21 & 78.19 & 35.30 & 79.81 & 34.55 & 83.07 & 35.56 & 78.94 & 35.16 & 80.00 \\    
                  FedDG~\cite{liu2021feddg} & 34.86 & 79.54 & 34.79 & 80.91 & 33.15 & 80.31 & 34.49 & 77.42 & 34.32 & 79.55\\       
                                    \rowcolor{lightgray}
    FedKD~\cite{wu2022communication}& 36.82 & 91.60 & 34.27 & 86.37 & 32.49 & 70.54 & 40.73 & 95.90 & 36.08 & 86.10 \\     \hline
    \multicolumn{11}{c}{Personalized FL Methods} 
\\
\hline
                  FedPer~\cite{arivazhagan2019federated} & 36.86 & 88.58 & 35.46 & 87.02 & 31.96 & 71.64 & 38.56 & 90.90 & 35.71 & 84.53\\
                                    \rowcolor{lightgray}
                  FedBN~\cite{li2021fedbn} & 37.84 & 89.95 & 36.16 & 88.46 & 33.67 & 81.63 & 40.21 & 94.89 & 36.97 & 88.73\\
                  FedMRI~\cite{feng2022specificity} & 31.12 & 82.42 & 31.95 & 79.95 & 31.72 & 71.96 & 32.04 & 81.63 & 31.71 & 78.99 \\  
                  \rowcolor{lightgray}
                  HyperFed~\cite{yang2023hypernetwork} & 36.18 & 89.60 & 33.47 & 82.49 & 31.94 & 65.52 & \textbf{41.70} & 94.85 & 35.90 & 83.12  \\
                  FedFDD~\cite{chen2024fedfdd} & 37.80 & 92.33 & 37.06 &91.77 & 33.71 & 80.26 & 40.53 & 95.12 & 37.27 & 89.87 \\
                  \hline 
                  \rowcolor{lightgray}
                   SCAN-PhysFed &  \textbf{40.77} & \textbf{97.69} & \textbf{38.77} & \textbf{95.75} & \textbf{34.38} & \textbf{81.80} & 40.27 & \textbf{96.92} & \textbf{38.55} & \textbf{93.04}\\
\hline
\end{tabular}
}
\end{table*}

\subsection{Generalization Evaluation}
\subsubsection{Backbone}
In addition to using RED-CNN as the imaging network in previous experiments, we also evaluate our method and other FL methods using an alternative imaging network, Uformer~\cite{wang2022uformer}, a transformer-based approach for LDCT imaging. This comparison further validates the generalization capability of the different FL methods.

The qualitative and quantitative results are shown in Figure~\ref{fig:uformer_sample} and Table~\ref{tb:uformer}, respectively. Consistent with previous experiments, our method achieves the best performance compared to other methods across most clients, with significantly higher average PSNR and SSIM values. This demonstrates the strong generalization capability of our method when integrated with different imaging networks. Among the other methods, FedFDD serves as a strong baseline; in a few clients, FedFDD's SSIM is slightly higher than ours, but our PSNR is markedly better, indicating that our method has superior detail preservation. Additionally, unlike FedFDD, which requires two imaging networks, our method achieves high performance without the need for an additional network. 
Similar to the previous experiment, the performance of SCAN-PhysFed shows better qualitative performance and less variation across different protocols, indicating that it maintains more consistent results regardless of the specific scanning protocols or imaging networks.


\begin{table}[!t]
\centering
\caption{Ablation on different components in SCAN-PhysFed.}
\label{tb:abla}
\resizebox{\columnwidth}{!}{%
\begin{tabular}{@{}cccc|cc@{}}
\hline
\multicolumn{4}{c|}{Component} & \multicolumn{2}{c}{Average} \\
Scanning & Anatomy & Paradigm & Loss & PSNR & SSIM \\  
\hline
\ding{55} & \ding{55} & Generic & $\mathcal{L}_\text{MSE}$ & \textcolor{gray}{34.03} & \textcolor{gray}{84.70} \\
\ding{51} & \ding{55} & Personalized & $\mathcal{L}_\text{MSE}$ & \textcolor{gray}{36.82} & \textcolor{gray}{88.78} \\
\ding{55} & \ding{51} & Personalized & $\mathcal{L}_\text{MSE}$ & \textcolor{gray}{38.04} & \textcolor{gray}{94.14} \\
\ding{51} & \ding{51} & Personalized & $\mathcal{L}_\text{MSE}$  & \textcolor{gray}{40.67} & \textcolor{gray}{97.27} \\
\ding{51} & \ding{51} & Personalized & $\mathcal{L}_\text{MSE}$ \& $\mathcal{L}_\text{orth}$ & \textbf{41.14} & \textbf{97.62} \\
\hline
\end{tabular}%
}
\vspace{-15pt}
\end{table}

\subsubsection{Unseen Protocol}
Existing FL methods in medical imaging fail to address how to ensure performance in unseen noise distributions from new clients. Here, we evaluate the effectiveness of the proposed PVQS across four unseen protocols; detailed protocol settings can be found in Appendix B. For generic FL methods, we directly evaluate the shared model on unseen clients. However, personalized methods mostly ignore how to operate in the unseen domain, so we average the performance of all local models to establish a vanilla baseline in this paper.

The results of different methods on unseen clients are presented in Table~\ref{tb:open}. We observe that the performance of generic FL methods on unseen clients shows only minor degradation compared to known clients, whereas personalized FL methods experience significant declines. This is because local optimization in personalized FL methods focuses heavily on each client’s specific data distribution, underfitting other potential distributions and leading to greater performance drops on unseen clients. In PVQS, we prevent the network from directly mapping the unseen protocol to modulate the imaging network, as this could cause the imaging network to collapse if the unseen protocol differs significantly from known protocols. Instead, we quantize the unseen protocol using a protocol codebook, allowing SCAN-PhysFed to leverage known knowledge to better adapt to unseen distributions.


\subsection{Ablation Study}

We evaluate the effectiveness of various components in SCAN-PhysFed, and the results are presented in Table~\ref{tb:abla}. Compared to the baseline results (which do not incorporate scanning and anatomy information), introducing either scanning or anatomy information significantly improves performance, highlighting the importance of integrating physical prior in LDCT imaging. This approach substantially reduces learning difficulty and improves the overall performance of the model.
Moreover, introducing both types of physical information simultaneously is not redundant but complementary, leading to further performance improvements. Building on this, we use $\mathcal{L}_\text{orth}$ to enhance the discriminative ability of the scanning code, making the imaging features more personalized and effectively mitigating the issue of data heterogeneity.
In the studies above, ``Personalized" indicates that the decoder is personalized for each client. An ablation study on the personalized components is provided in Appendix D.

\section{Conclusion}

To preserve privacy and address heterogeneity in CT imaging, we propose a dual-level, physics-informed approach that utilizes both scanning- and anatomy-level prompts to achieve personalized CT imaging with the assistance of an MLLM. This approach enables both patient- and scanning-level personalization, with extensive results demonstrating that incorporating physical prior significantly improves performance and effectively mitigates heterogeneity issues. Additionally, we propose a novel strategy, PVQS, to quantize protocol codes and maintain robust performance on unseen protocols. PVQS leverages known protocol codes to prevent latent collapse when the unseen protocol differs significantly from known protocols.
In future work, we plan to explore the development of a general FL framework for medical imaging that can support both CT and MRI denoising, which we believe holds promising potential.

%% file: sec/X_suppl.tex
\clearpage
\setcounter{page}{1}
\maketitlesupplementary

\setcounter{section}{0}
\setcounter{table}{0}
\setcounter{figure}{0}
\setcounter{equation}{0}

\renewcommand{\thefigure}{\Alph{figure}}
\renewcommand{\thesection}{\Alph{section}}
\renewcommand{\thetable}{\Alph{table}}


\begin{table*}[!t]
    \centering

    \caption{The Geometry Parameters and Dose Levels in Different Known Clients.}
    \resizebox{.8\textwidth}{!}{%
    \label{tb:close_set}
    \begin{tabular}{ccccccccc} 
    \hline
                        & Client \#1 & Client \#2 & Client \#3 & Client \#4 & Client \#5& Client \#6 & Client \#7 & Client \#8\\
      \hline


NV                                      & 1024    & 128 & 512 & 384 &  712 & 200 & 560 & 368        \\ 

NDB                                  & 512    & 768  & 768 & 600 & 720 & 730 & 755 & 500      \\ 

PL                                       & 0.66    & 0.78 & 1.00 & 1.40 & 0.60 & 0.88 & 1.20 & 1.00           \\ 

DBL                                 & 0.72          & 0.58  & 1.20 & 1.50 & 0.82 & 0.78 & 1.30 & 1.30 \\ 

DSR     & 250             & 350   & 500 & 350 & 300 & 350 & 300 & 350     \\ 

DDR & 250             & 300   & 400 & 300 & 350 & 280 & 400 & 350     \\ 

PN           & $1 \times 10^5$        &    $1\times 10^6$  &  $5\times 10^4$ & $1.25 \times 10^5$ &$1.3 \times 10^5$ & $ 0.9 \times 10^6$ & $4.5 \times 10^4$   & $1.45 \times 10^5$  \\ 
      \hline
      \end{tabular}
      }
      \vspace{-5pt}
\end{table*}

\begin{table*}[!t]
    \centering

   \caption{The Geometry Parameters and Dose Levels in Different Unseen Clients.}
   \vspace{-5pt}
\tiny
    \resizebox{.8\textwidth}{!}{%
    \label{tb:open_set}
    \begin{tabular}{cccccc} 
    \hline
                        & Unseen Client \#1 & Unseen Client \#2 & Unseen Client \#3 & Unseen Client \#4 \\
      \hline
NV      &  768 & 428 & 100 & 896 \\ 
NDB     &  550 & 590 & 768 & 730 \\ 
PL      &  0.57 & 1.10 & 0.50 & 0.70       \\ 
DBL     &  0.83 & 1.10 & 0.60 & 0.93 \\ 
DSR     &  200 & 350 & 200 & 250  \\ 
DDR     &  300 & 300 & 250 & 400  \\ 
PN      &  $1.3\times 10^5$ & $1.4\times 10^5$ &  $1.1\times 10^6$ &  $ 9 \times 10^4$   \\ 
      \hline
      \end{tabular}
      }
      \vspace{-10pt}
\end{table*}

\section{Implementation}

Differ with the previous works~\cite{feng2022specificity,chen2024fedfdd}, which only takes the image as the input ignoring the information contained in scanning. These paradigm can be formulated as:
\begin{equation}
    \mat{\hat{Y}} = D(E(\mat{X};\theta_E);\theta_D), \tag{A.1}
\end{equation}
where $E$ and $D$ denote the encoder and decoder, respectively.

Unlike previous approaches, our method simultaneously considers multiple physical factors influencing noise in the image domain, such as scanning protocols and anatomical structures. Different anatomical structures have varying attenuation coefficients, which result in diverse noise distributions~\cite{zaidi2003determination}, such as fat, blood, and bone.

To leverage anatomical information, we utilize the generation capabilities of the MLLM, miniGPT-Med~\cite{alkhaldi2024minigpt}, pretrained on multiple large-scale datasets. This method can effectively generate radiology reports for CT data. Specifically, we use the prompt, ``Please provide a radiology report of the CT slice." The feature extracted from the generated report by miniGPT-Med's text encoder is then used to personalize imaging features at the anatomy level.

To help readers in understanding the entire process of the proposed SCAN-PhysFed, we provide a pseudocode-style overview of our method in Algorithm~\ref{alg:Framwork}, where $\operatorname{PMB}(\cdot)$ denotes our proposed personalized modulation block.

  \begin{algorithm}[t]  
  \caption{Main steps of SCAN-PhysFed.}  
  \label{alg:Framwork}  
  \begin{algorithmic}[1]  
\Require $\mathcal{D} \triangleq U_{k \in K} \mathcal{D}^{k}$, data from $K$ institutions; $\mathcal{G} \triangleq \{G_1,...,G_K\}$; $C\triangleq \{c1,...,c_K\}$, the protocol set; $T$, the number of local training epoch; $Promp$, the prompt of MLLM, $\mathcal{M}$.
    \Ensure The global shared parameters $\theta_E$, $\theta_{H_s}$, $\theta_{H_a}$ for shared encoder $E$, scanning-informed and anatomy-informed hypernetworks $H_s$ and $H_a$, respectively. Additionaly, client-specified decoders $\{D_1,...,D_K\}$ for each of $K$ clients.
    \State \textbf{Server Executes:}
    \State Initialize $\theta_E$, $\theta_{H_s}$, $\theta_{H_a}$ and $\theta_D$ and deliver them to each client.
     \For{$t=1,2,...,N$}
        \For{$k=1,2,...,K$ \textbf{in parallel}} 
        \State send $\theta_E^t$, $\theta_{H_s}^t$, $\theta_{H_a}^t$ to $k$-th institution
        \State $\theta_E^{k,t}, \theta_{H_s}^{k,t}, \theta_{H_a}^{k,t} \gets $ \textbf{Local Train}($\theta_E^t, \theta_{H_s}^t, \theta_{H_a}^t$)
        \EndFor\\        \hspace{1.5em}$\theta_E^{t+1},\theta_{H_s}^{t+1},\theta_{H_a}^{t+1} \gets \sum_{k=1}^{K} \frac{\left|\mathcal{D}^{k}\right|}{|\mathcal{D}|} (\theta_E^{k,t},\theta_{H_s}^{k,t},\theta_{H_a}^{k,t})$
     \EndFor
    \State \textbf{Local Train:}
    \State $\theta_E^{t,1},\theta_{H_s}^{t,1},\theta_{H_a}^{t,1}  \gets \theta_E^{t},\theta_{H_s}^{t},\theta_{H_a}^{t}$
    \For{$e=1,2,...,E$}
        \For{$(\mat{X},\hat{\mat{Y}})$ in $\mathcal{D}^k$}
        \State $f_t \gets E(\mat{X})$
        \State $\alpha, \beta, c_k \gets H_s(g_k)$
        \State $f_{an} \gets H_a(\mathcal{M}(\mat{X},Promp))$
        \State $f_{per} \gets \operatorname{PMB}(f_{\mat{X}}, f_{an}, \alpha, \beta)$
        \State $\mat{Y} \gets D_k(f_{per})$
        \State $\mathcal{L}_{total} \gets \mathcal{L}_{\text{MSE}}(\mat{Y},\hat{\mat{Y}})+\tau \mathcal{L}_\text{orth}(c_k, C)$
        \State $\theta_E^{t,e},\theta_{H_s}^{t,e},\theta_{H_a}^{t,e} \gets \operatorname{BackProjection}$
        \EndFor
    \EndFor \\
    \Return $\theta_E^{t,E},\theta_{H_s}^{t,E},\theta_{H_a}^{t,E}$ to the server
  \end{algorithmic}  
\end{algorithm}


\section{Detailed Experimental Setup}

In CT imaging, the noise distribution is closely related to the scanning protocol, which primarily consists of seven parameters~\cite{wang2019machine}. These parameters include the number of views~(NV), the number of detector bins (NDB), pixel length (PL), detector bin length (DBL), the distance between the source and rotation center (DSR), the distance between the detector and rotation center (DDR), as well as the photon number of incident X-rays (PN). The units for DBL, DSR, and DDR are all in millimeters. 
To assist readers in re-simulating our data, we present the scanning protocols in Table~\ref{tb:close_set}. Besides, the scanning protocols of unseen clients can be found in Table~\ref{tb:open_set}. It can be observed that the protocol differences between known clients are quite different, and the protocols of unseen clients differ significantly from those of known clients. Following the previous simulation process~\cite{niu2014sparse}, we generate various LDCT data by introducing Poisson and electronic noise into the measured projection data to replicate low-dose conditions. The simulated funcion can be formulated as:
\begin{equation}
    \mat{Y}_{si} =\ln \frac{I_{0}}{\operatorname{Poisson}\left(I_{0} \exp (-\hat{\mat{Y}}_{si})\right)+\operatorname{Normal}\left(0, \sigma_{e}^{2}\right)}, \tag{A.2}
\end{equation}
where $\mat{Y}_{si}$ and $\hat{\mat{Y}}_{si}$ denote the clean projection and noised projection, respectively. $\sigma_{2}^2$ represents the electronic noise variance, which is set to 10 in this paper following~\cite{niu2014sparse}.

Since the parameters within $g$ have different value ranges, we normalized all the parameters in $g$ as:
\begin{equation}
    \hat{g^i_{j}}=\frac{g^i_{j}-\min \left(g_{j}\right)}{\max \left(g_{j}\right)-\min \left(g_{j}\right)}, \tag{A.3}
\end{equation}
where $g^i_j$ is $j$-th element in $i$-th protocol vector $g$. $\hat{g^i_j}$ denotes the normalized feature. $\max(g_j)$ and $\min(g_j)$ represent the maximum and minimum values of the $j$-th element across all $g$ of different strategies.


\begin{figure*}[t]
  \centering
   \includegraphics[width=.9\linewidth]{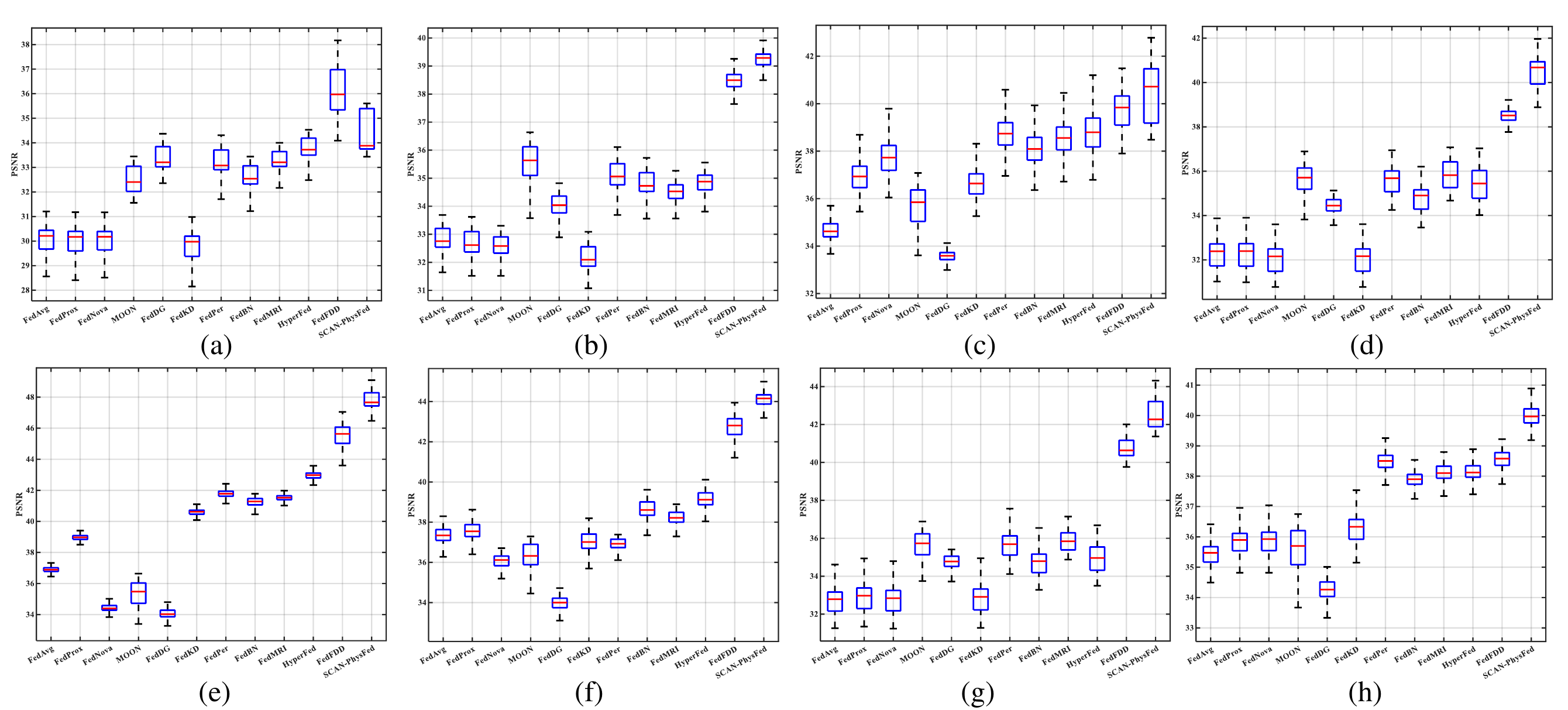}
   \vspace{-5pt}
   \caption{Boxplots of PSNR in all clients based on different FL methods. (a)-(h) indicate the PSNR boxplots of Client \#1 to Client \#8, respectively.}
   \label{fig:box_psnr}
\end{figure*}

\begin{figure*}[t]
  \centering
   \includegraphics[width=.9\linewidth]{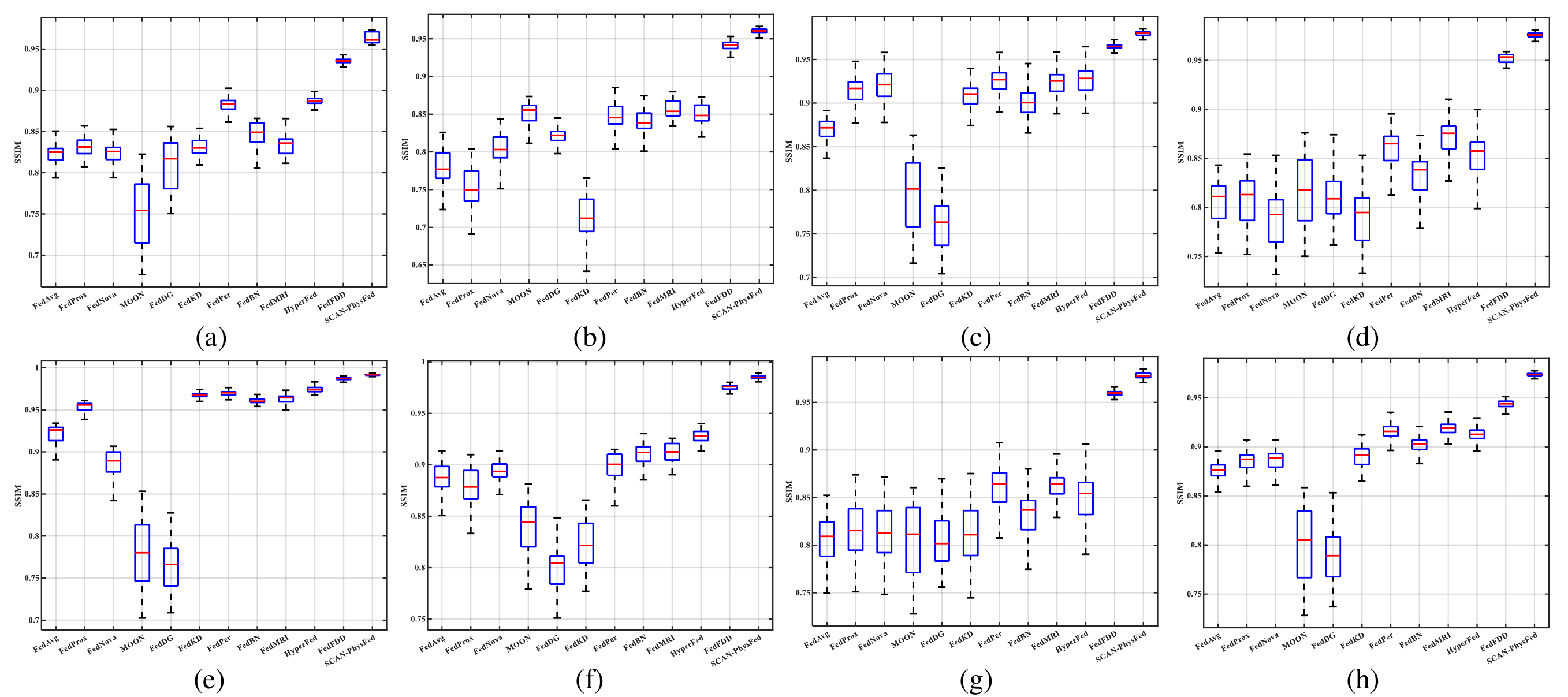}
   \vspace{-5pt}
   \caption{Boxplots of SSIM in all clients based on different FL methods. (a)-(h) indicate the SSIM boxplots of Client \#1 to Client \#8, respectively.}
   \label{fig:box_ssim}
\end{figure*}

\section{Stability Experiment}
Figures~\ref{fig:box_psnr} and~\ref{fig:box_ssim} illustrate the boxplots of both PSNR and SSIM for different FL methods in different clients. As seen in the figures, our SCAN-PhysFed not only achieves a consistently high median but also exhibits a narrower interquartile range across all clients, indicating low variability and minimal outliers. This stability underscores the robustness of my method in comparison to others. In contrast, other methods often show wider interquartile ranges and more extreme outliers, reflecting considerable performance fluctuations among clients. Specifically, generic FL methods can maintain reasonable performance in certain clients, yet they often collapse in clients with different noise distributions. While personalized FL methods partially address this issue, they still exhibit poor performance in some clients because they lack sufficient knowledge to effectively personalize the imaging process.

Our method, SCAN-PhysFed, adopts a physics-driven approach, leveraging physical scanning parameters and anatomy priors to personalize the imaging process. This approach effectively reduces learning complexity and enhances overall model performance. The results strongly support this, with SCAN-PhysFed displaying consistent boxplot characteristics across all subplots, underscoring its robustness and reliability in maintaining steady performance with minimal variation.



\section{Personalization Experiment}

Additionally, we evaluate the performance with different personalized components. 
In these experiments, we use RED-CNN as the imaging backbone and optimize the network using only MSE loss, with other settings consistent with previous experiments. Attempts to only personalize the encoder resulted in non-converging training, suggesting that the encoder may not be suitable as a personalized component. Meanwhile, it is challenging to use a shared decoder to project varying imaging features into the image domain.  Furthermore, since anatomy prior is independent of the scanning protocol and remains similar across different patients, we chose to leverage data from multiple clients to train a shared, robust anatomy feature extractor.

Therefore, in this study, we focus on personalizing the decoder and the scanning-informed hypernetwork. Results in Table~\ref{tb:pesona} show minimal performance differences between these methods, indicating that both personalized decoder and hypernetwork approaches yield satisfactory results. However, to handle unseen clients, our PVQS strategy requires quantizing the scanning code using a protocol codebook, which necessitates extracting these scanning codes with a standardized extractor. As a result, we adopt client-specific decoders in this paper to project personalized features into the image domain.

\section{Limitation}

Although the proposed SCAN-PhysFed demonstrates satisfactory generalizability across different protocols, body parts, and imaging networks, there are still areas for improvement in future work. \textbf{\textit{i)}} This method is closely aligned with the physical principles of CT imaging; therefore, adapting it to other imaging modalities would require adjustments based on the specific physical characteristics of those modalities. A promising direction for future work is to extend this approach to other medical imaging modalities, such as MRI or ultrasound, by incorporating relevant modality-specific physical knowledge. \textbf{\textit{ii)}} While introducing anatomical information significantly enhances performance, it also raises potential security concerns. For instance, an attacker could compromise the MLLM and inject a backdoor into the feature representation, potentially deceiving downstream tasks and leading to inaccurate or manipulated results. This underscores the need for robust security measures to protect the model against such vulnerabilities.

\begin{table}[!t]
\centering
\caption{The Quantitative Results of PSNR and SSIM for the Post-processing Task.}
\label{tb:pesona}
\resizebox{\columnwidth}{!}{%
\begin{tabular}{@{}c|cc@{}}
\hline
\multicolumn{1}{c|}{} & \multicolumn{2}{c}{Average} \\
Personalized Component   & PSNR & SSIM \\  
\hline
Decoder $E$ & 40.67 & 97.27 \\
Scanning-Informed Hypernetwork $H_s$ & 40.38 & 97.22 \\
Decoder $E$ \& Scanning-Informed Hypernetwork $H_s$ &  40.63 & 97.25\\
\hline
\end{tabular}%
}
\vspace{-5pt}
\end{table}

%